\documentclass[aps, prl,twocolumn,superscriptaddress, reprint, nofootinbib, showpacs]{revtex4-2}
\usepackage{graphicx}% Include figure files
\usepackage{dcolumn}% Align table columns on decimal point
\usepackage{textgreek}
\usepackage{adjustbox}
\usepackage{textcomp}

\def\KtCA{K\textsubscript{2}Cr\textsubscript{3}As\textsubscript{3}}
\def\KtMA{K\textsubscript{2}Mo\textsubscript{3}As\textsubscript{3}}

\def\ACA{\textit{A}\textsubscript{2}Cr\textsubscript{3}As\textsubscript{3}}
\def\AoCA{\textit{A}Cr\textsubscript{3}As\textsubscript{3}}

\def\SRO{Sr\textsubscript{2}RuO\textsubscript{4}}

\def\iA{\text{\AA}\textsuperscript{-1}}

\def\Psmt{$P\overline{6}m2$}

\def\Tc{$T_C$}

\begin{document}

\title{Gapless spin-excitations in the superconducting state of a quasi-one-dimensional spin-triplet superconductor}
 
%Authors
\author{Keith M. Taddei}
\email[corresponding author ]{taddeikm@ornl.gov}
\affiliation{Neutron Scattering Division, Oak Ridge National Laboratory, Oak Ridge, TN, 37831, USA}
\author{Bing-Hua Lei}
\affiliation{Department of Physics and Astronomy, University of Missouri, Missouri 65211, USA}
\author{Michael A. Susner} 
\affiliation{Materials and Manufacturing Directorate, Air Force Research Laboratory, Wright-Patterson Air Force Base, OH,45433, USA }
\author{Hui-Fei Zhai}
\affiliation{University of Dallas, Richardson, TX, USA}
\author{Thomas J. Bullard}
\affiliation{Aerospace Systems Directorate, Air Force Research Laboratory, Wright-Patterson Air Force Base, OH, 45433 USA }
\affiliation{UES, Inc., 4401 Dayton Xenia Road, Dayton, Ohio, 45432, USA}
\author{Liurukara D. Sanjeewa}
\affiliation{Materials Science Division, Oak Ridge National Laboratory, Oak Ridge, TN, 37831, USA}
\author{Qiang Zheng}
\affiliation{Materials Science Division, Oak Ridge National Laboratory, Oak Ridge, TN, 37831, USA}
\author{Athena S. Sefat}
\affiliation{Materials Science Division, Oak Ridge National Laboratory, Oak Ridge, TN, 37831, USA}
\author{Songxue Chi}
\affiliation{Neutron Scattering Division, Oak Ridge National Laboratory, Oak Ridge, TN, 37831, USA}
\author{Clarina dela Cruz}
\affiliation{Neutron Scattering Division, Oak Ridge National Laboratory, Oak Ridge, TN, 37831, USA}
\author{David J. Singh}
\email{singhdj@missouri.edu}
\affiliation{Department of Physics and Astronomy, University of Missouri, Missouri 65211, USA}
\author{Bing Lv} 
\email{blv@utdallas.edu}
\affiliation{University of Dallas, Richardson, TX, USA}

\date{\today}

\begin{abstract}

Majorana zero modes form as intrinsic defects in an odd-orbital one-dimensional superconductor thus motivating the search for such materials in the pursuit of Majorana physics. Here, we present combined experimental results and first principles calculations which suggest that quasi-one-dimensional \KtCA\  may be such a superconductor. Using inelastic neutron scattering we probe the dynamic spin-susceptibilities of \KtCA\ and \KtMA\ and show the presence of antiferromagnetic spin-fluctuations in both compounds. Below the superconducting transition, these fluctuations gap in \KtMA\ but not in \KtCA . Using first principles calculations, we show that these fluctuations likely arise from nesting on one dimensional features of the Fermi surface. Considering these results we propose that while \KtMA\ is a conventional superconductor, \KtCA\ is likely a spin-triplet, and consequently, topological superconductor.  
 
\end{abstract}

\maketitle

%Despite the decades long anticipation of quantum computers, only recently has their implementation become feasible with the realization of multi-qubit processors which has in turn driven a growing excitement for their myriad applications\cite{Feynman1982,Sau2017,Materise2018}. Unfortunately, the current technologies behind the quantum computer's defining qubit have proven incredibly difficult to scale creating a roadblock and consequentially a deep need for new phenomenon on which to base the qubit \cite{Arute2019,Ballance2016,Devoret2013,Geller2007,Petersson2010}. This has led to an effort to use the robustness and the quantum-entanglement of topological states as a new basis for qubit design \cite{Kitaev2003,Nayak2008,Stern2013,Leon2021}. Of the potential candidates the Majorana zero mode (MZM) has stood out as one of the most promising due to its non-abelian anyon statistics which are suited for braiding, its building off existing research in superconductivity, and its potential to be physically manipulated as is necessary to perform actual operations \cite{Beenakker2013,Kitaev2001m,Vaitiekenas2020,Xiong2021,Alicea2011,Tutschku2020}. 

To realize scaleable quantum computers, new phenomena on which to base the qubit are needed - ones robust and with intrinsic entangled properties such as exists in certain topological phases \cite{Arute2019,Ballance2016,Devoret2013,Geller2007,Petersson2010,Kitaev2003,Nayak2008,Stern2013,Leon2021}. Of the potential candidates, the Majorana zero mode (MZM) is one of the most promising due to its non-abelian anyon statistics which are suited for braiding while also potentially allowing for physical manipulation as is necessary for computation \cite{Beenakker2013,Kitaev2001m,Vaitiekenas2020,Xiong2021,Alicea2011,Tutschku2020}. However, generating and observing MZMs has proven challenging with several potential routes in active pursuit such as: at the interface of a topological insulator and a superconductor (SC), in a SC with a topological band structure, or in a SC whose pair operator is its own conjugate - a \lq spin-less\rq\ or spin-triplet odd-orbital SC \cite{Chamon2017}. Indeed, this last case (when restricted to one dimension) is related to the toy-model first proposed by Kitaev to generate physically separated MZM  \cite{Kitaev2001m}.    

As a result, there is great interest in one-dimensional (1D) or quasi-1D (Q1D) systems which might exhibit spin-triplet SC (TSC) - especially those that exhibit both properties intrinsically. However, such materials are extraordinarily rare with few compounds showing either property and still fewer with both. Nonetheless, several candidate materials have been found (including the Bechgaard salts, purple bronze, and even possibly \SRO )\cite{Brown2015,Cho2015, Firmo2013}. More recently, the discovery of the Q1D potential TSC $A_{n}$H$_{(2-n)x}TM_3$As$_3$ (with $A$ = Na,K,Rb or Cs, $TM$ = Cr or Mo and $n$ = 1 or 2) family has provided another route to realize these exotic physics   \cite{Bao2015a,Bao2015,Xiang2019,Tang2015r,Tang2015c,Tang2015r1,Mu2018c,Mu2018b,Liu2017,Taddei2019}. 

%. This can be achieved via engineering heteromaterials with ferromagnets in close proximity to a superconductor \cite{Lutchyn2018,Manna2020}. Another route is to find materials that are both inherently Q1D due to their crystal structures and exhibit spin-triplet superconductivity \cite{Kaladzhyan2017,Ray2021}. Such systems may not only lead to simpler potential device architectures but also may host higher order topological phases \cite{Ray2021}. 

The $A_{n}$H$_{(2-n)x}TM_3$As$_3$ materials exhibit numerous novel properties, several of which evince TSC. These materials  crystallize with a Q1D structural motif of $TM_3$As$_3$ tubes which give rise to strongly Q1D features such as Luttinger-liquid physics, Q1D Fermi surfaces (FS) and highly anisotropic transport \cite{Bao2015a,Bao2015,Cao2017,Mu2017,Mu2018b,Noce2020,Watson2017}. Enticingly, their SC state appears to be unconventional with an unexpectedly high upper critical field, potential nodes in the SC gap, and a proximity to a quantum critical point with possible suggestions of TSC due to a spontaneous magnetization below the SC transition (\Tc ), an angular dependent upper critical field, ferromagnetic (FM) fluctuations, a \Tc\ suppressed by non-magnetic impurities and widespread findings of a leading TSC instability from theory \cite{Balakirev2015, Pang2015,Luo2019,Adroja2015, Liu2016, Zuo2017, Zhi2015,Wu2015a,Zhong2015, Zhang2016, Xu2020,Wu2019,Cuono2021}.  

However, the symmetry of the SC state, and thus the prospects for hosting MZM, remains disputed. While some studies support TSC, others have suggested a spin-singlet state. These investigations report anti-FM (AFM) instabilities, a proximity to a spin-glass state, a $s^{\pm}$ gap symmetry, and even claims of standard electron-phonon (\textit{e-p}) coupling \cite{Taddei2017t, Taddei2019, Cao2015,Subedi2015}.  Recently, it was proposed that the K$_2TM_3$As$_3$ family may straddle a boundary between an unconventional SC in \KtCA\ (\Tc $\sim6$ K) and a multi-gap conventional SC in \KtMA\ (\Tc $\sim10$ K), perhaps giving some guidance to understand the disparate reported features\cite{Lei2021}.   
 
In this Letter, we assess the possibility of TSC in \KtCA\ through a careful study of the dynamic spin susceptibilities of \KtCA\ and \KtMA\ using both experimental probes and first principles calculations. To start, inelastic neutron scattering (INS) experiments reveal spin-fluctuations (SF) in both compounds above \Tc\ which are consistent with an incipient AFM order. Below \Tc , we find that for \KtMA\ a resonanceless spin-gap opens while in \KtCA\ no gap is observed implying a difference in the compounds' SC states. Performing first principles calculations, we find that the AFM SF can be explained by FS nesting on one dimensional FSs. Consequently, we suggest that \KtMA\ is an \textit{e-p} SC whose low energy SF are suppressed due to the opening of SC gaps on all FSs.  Contrastingly, the lack of a spin-gap in \KtCA\ indicates that neither the AFM SF nor the associated FSs participate in SC leaving a single remaining FS which much be SC and is favorable to FM SF driven TSC. These conclusions support the scenario of FM driven TSC in \KtCA\ and indicate its potential for hosting topological SC.    

%To discriminate between superconducting models by studying the fluctuations in the superconducting state for both \KtCA\ and \KtMA . In \KtCA\ we find no spin gap thus indicating that the relevant Fermi surface is not gapped in the superconducting state. On the other hand, in \KtMA\ we find a clear gap opens below \Tc\ indicating the nested Fermi surface's gapping in the superconducting state. Using density functional theory, we identify which sheets nest to give rise to the AFM fluctuations as the three dimensional $\gamma$ sheet. From these results, we suggest that in \KtCA\ superconductivity lives on the Q1D Fermi Surfaces and is not induced by the AFM spin-fluctuations indicating a likely TSC state. On the other hand, the gapping of the spin-fluctuations in \KtMA\ is consistent with recent predictions of a multi-gap e-p superconducting mechansim where superconductivity gaps all of the Fermi surfaces. These results strongly suggest that \KtCA\ is a TSC Q1D superconducting thus encouraging a search for MZM. 
%Low dimensionality, and quantum physics - Q1D, superconductivity, topological/spin-triplet, and luttinger-liquid. 

%ATMAs superconductors, spin-triplet? unconventional? properties. Fluctuations, no-long range order. Possible instabilities. Arguments for exotic physics and spin-triplet/topological. 

%David's paper and the difference between the Fermi surfaces. Motivation for current work - why compare KMA to KCA to learn something

\begin{figure}	
\includegraphics[width=\columnwidth]{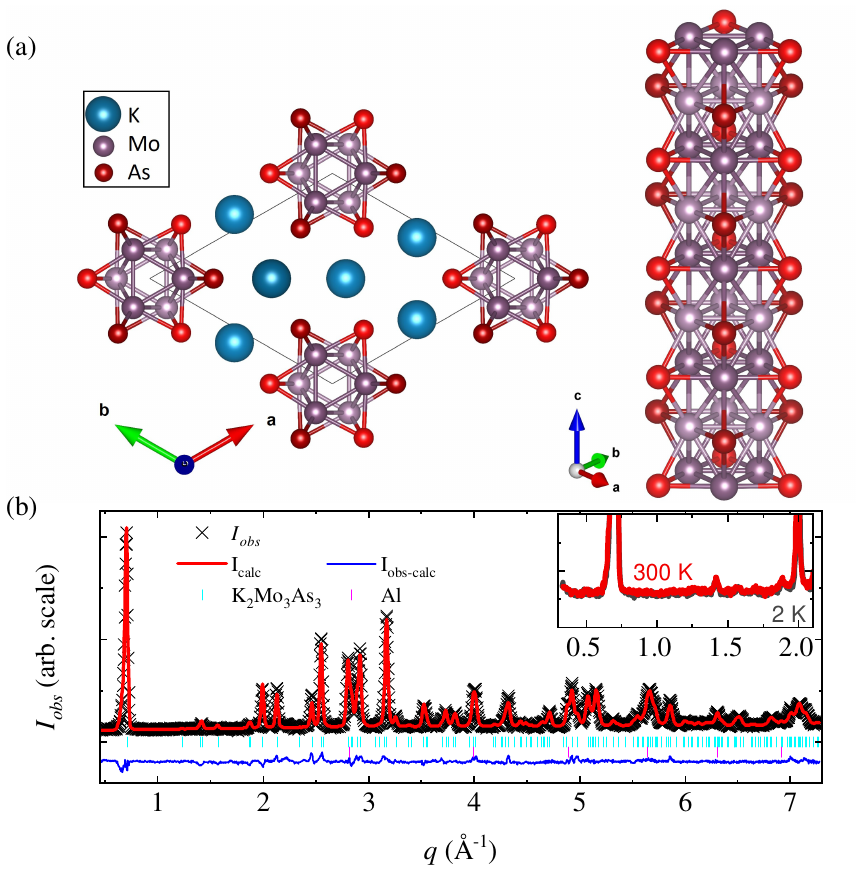}
	\caption{\label{fig:one}  (a) Crystal structure of \KtMA\ viewed along \textit{c} and of the isolated tube motif. (b) Neutron powder diffraction pattern and best model fit for data collected at 300 K. Inset of panel (b) shows a comparison of the low $q$ region of data collected at 300 and 2 K.}	
\end{figure}

Large powder samples of \KtCA\ and \KtMA\ were synthesized using methods reported previously to obtain $\sim 10$ g of materials per compound (see the supplemental materials (SM) for details) \cite{Taddei2017t,Bao2015a,Mu2018b}.  Neutron powder diffraction (NPD) measurements were performed on the HB-2A diffractometer of Oak Ridge National Laboratory's (ORNL) High Flux Isotope  Reactor (HFIR) \cite{Calder2018}. The resulting diffraction data were analyzed using the Rietveld method as implemented in the FullProf software Suite \cite{Rodriguez-Carvajal1993}. INS was performed on the HB-3 and CTAX triple axis spectrometers of HFIR using fixed analyzer energies of 14.7 and 5 meV respectively.  To calculate the electronic band structure, Density Functional Theory (DFT) calculations were performed using the generalized gradient approximation of Perdew, Burke and Ernzerhof (PBE) and the general potential linearized augmented planewave method as implemented in the WIEN2k code \cite{Perdew1996,Singh2006,Blaha2020}. 

In fig.~\ref{fig:one} (a) we show the crystal structure of \KtMA\ (space group \Psmt ) which exhibits a unique Q1D structural motif comprised of two inequivalent, alternating, coaxial layers of Mo (and As) triangles. In fig.~\ref{fig:one}(b) we show a diffraction pattern of \KtMA\ collected at 300 K together with a simulated pattern from our best-fit model (we note that we found no impurity phase in our diffraction data indicating the high quality of our sample). In the inset of fig.~\ref{fig:one}(b) we show a comparison of NPD patterns collected at 300 and 2 K demonstrating a lack of any significant changes which might be associated with the onset of magnetic order indicating that \KtMA\ (as \KtCA) has no long-range magnetic order (see the SM for more discussion) \cite{SM}.

%To start our comparison of \KtCA\ and \KtMA\ we first review the nuclear structure and consider the possibility of long-range magnetism in \KtMA\ (see ref.~\onlinecite{Taddei2017t} for similar work on \KtCA ). In fig.~\ref{fig:one} (a) we show the crystal structure of \KtMA\ (space group \Psmt ) which exhibits a unique quasi-one-dimensional (Q1D) structural motif comprised of two inequivalent, alternating and coaxial layers of Mo (and As) triangles with 60\degrees\ rotations between creating a hexagram projection when viewed along the \textit{c}-axis. With the hexagonal symmetry, these so-called double-walled sub-nano tubes (for the concentric Mo and As tubes) form a triangular lattice with the K$^+$ ion filling the channels between the tubes. 

%In fig.~\ref{fig:one}(b) we show a diffraction pattern of \KtMA\ collected at 300 K together with a simulated pattern from our best-fit model (we note that we found no impurity phase in our diffraction data indicating the high quality of our sample). In the inset of fig.~\ref{fig:one}(b) we show a comparison of NPD patterns collected at 300 and 2 K demonstrating a lack of any significant changes which might be associated with the onset of magnetic order. We therefore conclude that \KtMA\ (as \KtCA) has no long-range magnetic order. For more discussion of the structure see the SM \cite{SM}.

Previously, \KtCA\ was shown to have AFM SF arising from incipient $\textbf{k}=(0,0,\frac{1}{2})$ type order. This was revealed as a column of scattering in the dynamic structure factor $S(q,\Delta E)$ (as probed via INS) which is proportional to the imaginary component of the spin-susceptibility \cite{Taddei2017t,Zaliznyak2004,Whitt2021}. Such fluctuations and their temperature dependence offer significant insights to the SC state. Therefore, we now turn to similar experiments performed on \KtMA\ and expand on our previous work on \KtCA\ to compare their respective $S(q,\Delta E)$.

\begin{figure}	
\includegraphics[width=\columnwidth]{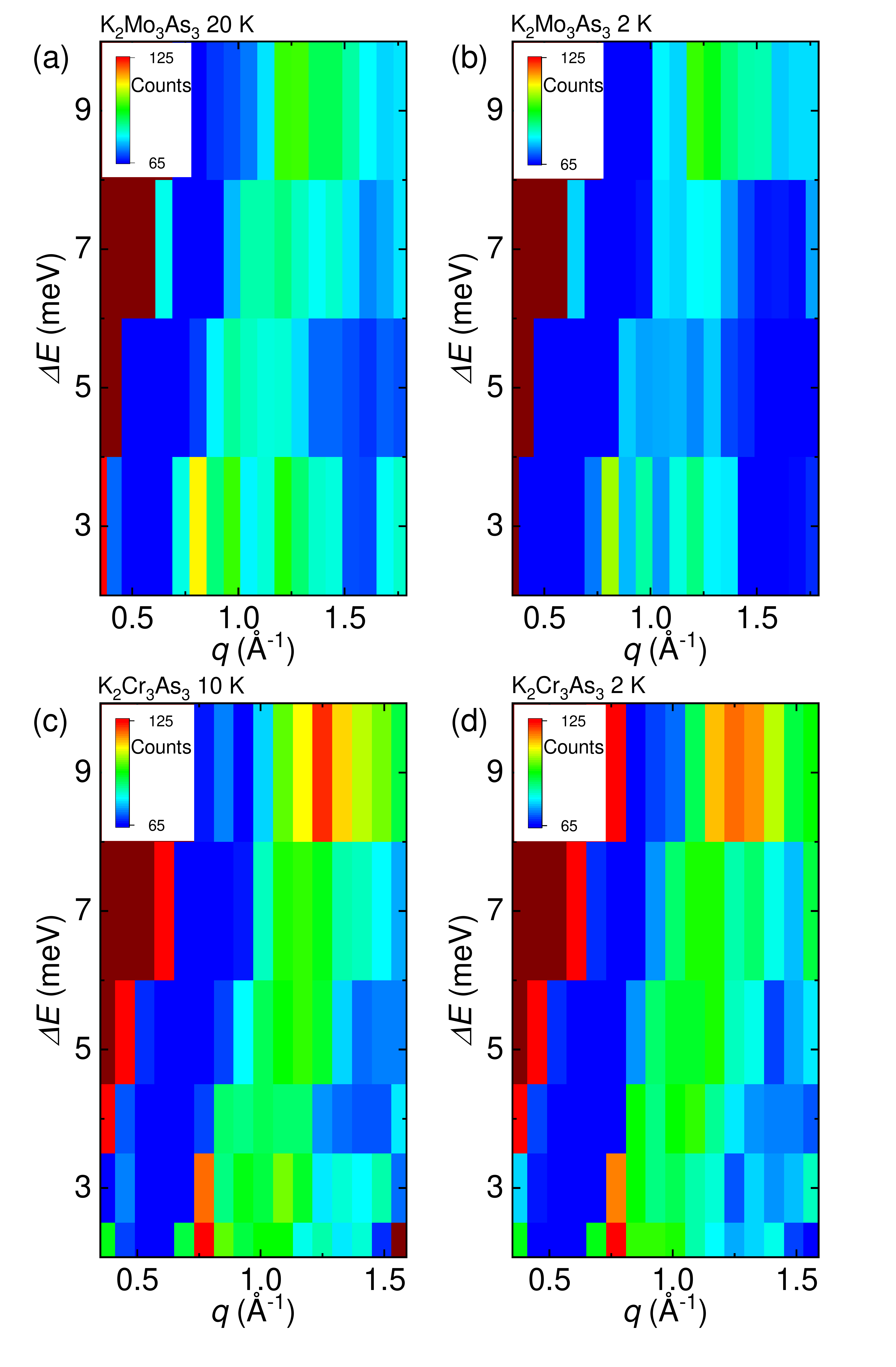}
	\caption{\label{fig:two}  Inelastic neutron scattering spectrograms for \KtMA\ at (a) 20 K and (b) 2 K and for \KtCA\ at (c) 10 K and (d) 2 K. Intensity is in units of detector counts normalized to monitor counts. We note that the data showed in panel (c) includes data from ref.~\onlinecite{Taddei2017t} but with additional counting statistics.}	
\end{figure}

Fig.~\ref{fig:two}(a) and (c) show the $S(q,\Delta E)$ of \KtMA\ and \KtCA\ collected at 20 and 10 K respectively (i.e. above either compound's \Tc ). Here we focus on the low \textit{q} and low $\Delta E$ region which is typically featureless at these temperatures for non-magnetic materials. However, for both materials a column of scattering is seen arising from $\sim$0.75 \iA . Such a signal is often indicative of incipient magnetic order caused by SF with a $q$ characteristic of the incipient ordering vector  \cite{Zhi2015,Zhi2016,Rahn2015,Taylor2011,Wang2016, Whitt2021}.

Qualitatively, the signal observed in \KtMA\ is similar to that of \KtCA . Fitting the constant $\Delta E$ cuts of the two datasets with Gaussian functions, we find a slight shift in the position of the feature to lower $q$ by $\sim$ 0.1 \iA\ in \KtMA\ compared to \KtCA\ which is consistent with the larger $c$ axis (see SM for details) \cite{SM}. Furthermore, the dispersion of the two signals is very similar (albeit they are both skewed due to the instrument's resolution function). On the other hand, the fits reveal that the column in \KtMA\ is broader in $q$ by $\sim 20\%$ and also is $\sim 30\%$ weaker (though this is more difficult to reliably quantify between samples) which may indicate the fluctuations are shorter-ranged and a smaller fluctuating moment is present in \KtMA\ - both of which have been suggested from prior DFT treatments \cite{Lei2021}. Due to these considerations, we attribute the origin of this signal to similar causes as in \KtCA . 

With the origin identified, we now turn to the temperature dependencies across \Tc . Fig.~\ref{fig:two}(b) and (d) show the same region of $S(q,\Delta E)$ measured below \Tc\ at 2 K for both samples. Here a distinction between the two emerges. For \KtCA\ the spectrograph looks qualitatively identical to the 20 K data set. In particular, no gap opens in the fluctuations despite the onset of SC. On the other hand, in \KtMA\ (fig.~\ref{fig:two}(b)) there is a clear change in the column where the signal for $\Delta E < 7$ meV loses intensity. This observation is consistent with the opening of a SC gap which inhibits fluctuations below $2\Delta$ (i.e. the energy required to break a Cooper pair).

\begin{figure}	
\includegraphics[width=\columnwidth]{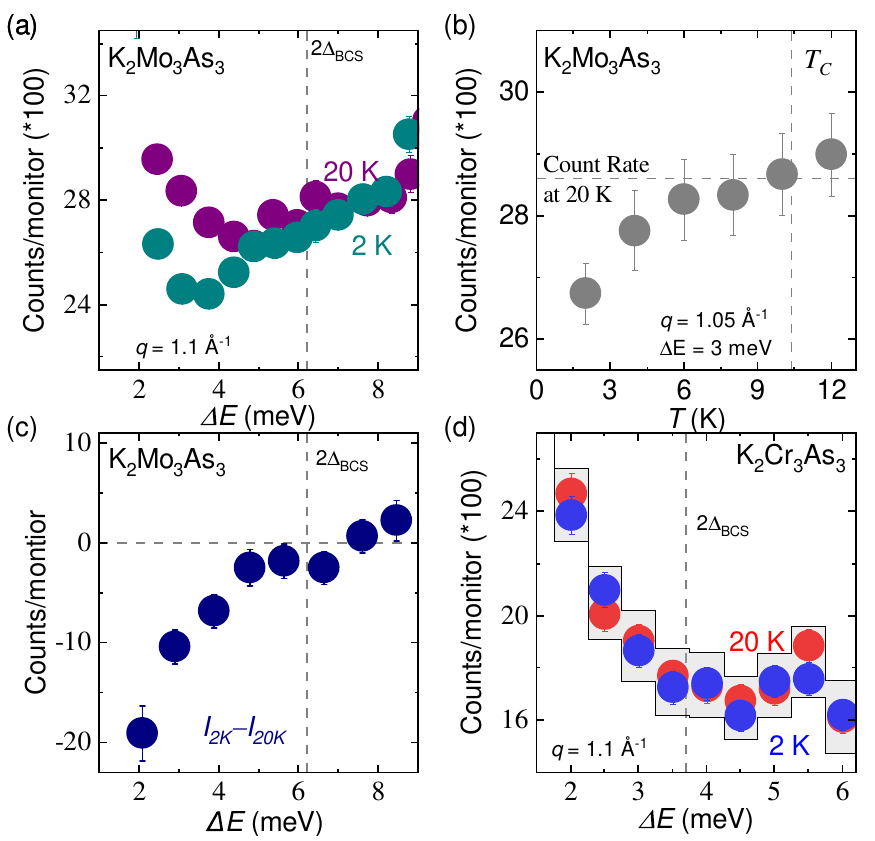}
	\caption{\label{fig:three}  (a) Comparison of scattering intensity of \KtMA\ for constant \textit{q} scans along the column collected at 20 and 2 K. (b) Temperature dependence of the low energy region of the \KtMA\ column with the 20 K count rate and \Tc\ denoted by horizontal and vertical dotted lines. (c) Difference curve for the 20 and 2 K scans of \KtMA\ shown in panel (a). (d) Similar comparison of 20 K and 2 K scans for \KtCA\ with the an envelope denoting the size of a gap expected for a signal similar to that observed in panel (a).}	
\end{figure}

To characterize this feature more carefully, constant $q$ scans were taken at $q \sim\ 1.1$\iA\ above and below \Tc\ for both samples (fig.~\ref{fig:three}). For \KtMA (fig.~\ref{fig:three}(a)), the gap becomes clear. While the 20 K data exhibit a constant increase in intensity below 5 meV (as the elastic line is approached), the 2 K data exhibit a qualitatively different behavior, dropping in intensity by $\sim 20\%$ below $\sim 5$ meV. Using the weak coupling Bardeen Cooper Schieffer gap approximation (i.e. $\Delta (T=0) = \frac{7}{2}k_BT_C$) we estimate $2\Delta$ as 6.2 meV which is consistent with our observed gap (a similar estimate is obtained using the empirical formula of $\omega_0 = 4.3 k_BT_c$ with $\omega_0$ being the energy of the spin-gap in the SC state)\cite{Christianson2008}. In fig.~\ref{fig:three}(c) we show a difference curve between the 20 and 2 K data to remove background effects. Here, the gap is seen to open below $~\sim 5$ meV and progressively widen to the lowest measured temperature of 2 K. We note that the shape of this curve may corroborate nodal or nodeless gap functions however, we do not believe our data are sufficient to allow such an analysis. We further associate this gap with \Tc\ by measuring the intensity at 1.05 \iA\ and 3 meV as a function of temperature (fig.~\ref{fig:three}(b)) which shows the gap to close at $\sim 6$ K. This is a little below \Tc\ (10.4 K); however, the gap itself is a function of $T$ and so should become smaller than the certainty of our measurements before \Tc\ is exceeded. 

Turning to \KtCA , we see discretely different behavior in the low energy spectrum (fig.~\ref{fig:three}(d)). Here, no obvious gap in the fluctuations is seen in the 2 K data. If the gap size is estimated as before, $2\Delta \sim 3.7$ meV and  $ \omega_0 \sim 2.2$ meV, both of which are within the limits of our energy resolution ($\sim 1$ meV). For comparison, in fig.~\ref{fig:three}(d) we plot an envelope showing the range equivalent to the percent change of the signal seen in \KtMA , demonstrating that, within our statistics, a similar decrease in intensity should be observable if present. Additional measurements were taken using a cold neutron triple-axis spectrometer to access lower energy transfers ( $<1$ meV) and no gap was observed (see SM)\cite{SM}. Consequently, we take this observation to be a strong indication that no spin-gap opens in the SC state of \KtCA .  

Such an observation has significant implications for the nature of SC in these systems as well as how it evolves between the two materials \cite{Lei2021}. That the SF in \KtCA\ do not respond strongly to SC requires an explaination - naively SC should open a gap.  Furthermore, though a spin-gap with an accompanying resonance has become a hallmark of unconventional SCs near magnetic order, here we see no evidence of a resonance above the gap in \KtMA\ and no resonance or gap in \KtCA\ undermining SF possible role in SC \cite{Stewart2017}. To help interpret these observations we note that if the SF can be associated with specific features of the FSs then the presence (or absence) of a gap in those SF will directly correspond to the presence (or absence) of a gap on the associated FS. In a system such as \KtCA , where different FSs have different SC instabilities, such information can be key in determining the symmetry of the SC state  \cite{Mason1992,Dahm1998,Dai1999,Christianson2008,Chi2009,Osborn2009,Park2011,Qureshi2012,Scalapino2012,Zhang2013,Tranquada2014,Wang2016n,Xie2018,Kunkemoller2017}.

\begin{figure}	
\includegraphics[width=\columnwidth]{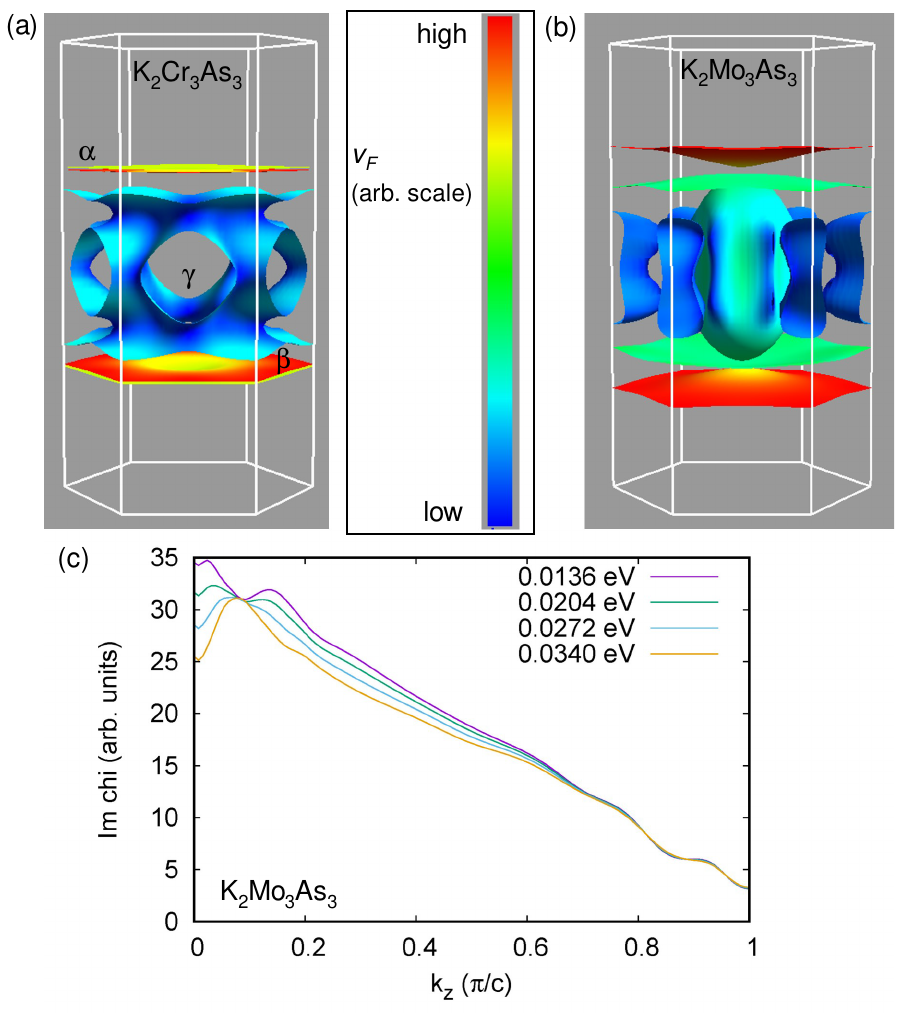}
	\caption{\label{fig:four}  Calculated Fermi surfaces of (a) \KtCA\ (undistorted) and (b) \KtMA\ . In (a) and (b) the calculated Fermi velocity is shown as a function of position on the Fermi surface via the color scale with blue indicating low relative velocities and red indicating higher velocities. (c) Imaginary component of the calculated Lindhard susceptibility of \KtMA\ plotted for several energies near the Fermi energy (with $E_F = 0$ eV).  }	
\end{figure}

To elucidate the origin of the SF, we consider the FSs of \KtMA\ and (undistorted) \KtCA\ (shown in fig.~\ref{fig:four} (a) and (b)) as determined by DFT calculations. Here we use undistorted \KtCA\ due to ambiguity in the exact symmetry of the distorted structure \cite{Taddei2018,Xing2019}.  As reported, these two compounds have similar FSs, consisting of two Q1D $\alpha $ and $\beta$ sheets and one large 3D $\gamma$ sheet \cite{Lei2021,Xu2020,Wu2015a,Alemany2015,Jiang2015,Yang2019}. Given the large sheet-like features of the FSs, nesting vectors have long been proposed as possible between both the upper and lower $\alpha$ and $\beta$ sheets as well as between the top and bottom surfaces of the $\gamma$ sheet any of which may lead to spin- or charge-density wave type orders with such a mechanism being proposed for the SF observed in \KtCA \cite{Lei2021,Subedi2015,Zhi2015,Zhang2019,Xing2019,Taddei2017t}.

With the potential for FS nesting established, we next calculate the Fermi velocities ($v_F$) throughout the FSs of both compounds to predict the strength of electron correlations on the different surfaces (as shown in the color scale on fig.~\ref{fig:four} (a) and (b)).  These calculations reveal two important features: for both compounds the large 3D $\gamma$ sheet has significantly lower $v_F$ indicating stronger correlated electron physics on this sheet. Additionally, $v_F$ is in general larger in \KtMA\ suggesting weaker electron correlations. These results suggest both that magnetic interactions should be stronger on the $\gamma$ sheet and stronger in \KtCA\ than in \KtMA\ in general. % This former point provides a significant insight, suggesting that phenomenon such as an incipient FS nesting driven SDW should exist on the more strongly correlated $\gamma $ sheet.       

%Doing so allows us to predict the relative strength of electron correlations on the different features of the FSs with high $v_F$ corresponding to relatively weak correlations and low $v_F$ indicating stronger correlations.

With the Fermiology indicating potential nesting, we turn to calculations of the imaginary component of the Lindhard susceptibility (which corresponds to the dynamic spin susceptibility) projected along $k_z$ for \KtMA\ (fig.~\ref{fig:four}(c)) (for this purpose we used a very dense grid of \textit{k}-points and the constant matrix element approximation). Peaks in the Lindhard susceptibility have previously been reported in \ACA\ and \AoCA\ and attributed to SF with a nesting vector along $k_z$ consistent with intra-band scattering \cite{Lu1990,Wang2010b,Xu2020,Zhang2019}. Here, we most prominently see a large broad peak near the zone center which corresponds to the FM SF observed in nuclear magnetic resonance measurements \cite{Zhi2015,Zhi2016,Yang2015,Gosar2020}. However, at larger $k_z$, near the zone boundary, we see a second feature at $k_z \sim 0.9$ which closely corresponds to the $\textbf{k}=(0,0,\frac{1}{2})$ position for the AFM SF observed in INS. This peak is quite small which is consistent with it arising from nesting between the two high $v_F$ Q1D sheets. 

These insights from first principles allow us to better interpret the experimental results. They show that both \KtMA\ and \KtCA\ have similar potential nesting vectors involving the 1D FSs consistent with the observed column of SF. That the AFM SF do not gap in \KtCA\ indicates that neither the AFM SF nor the 1D FSs participate in SC. This is expected as symmetry considerations for AFM SF mediated spin-singlet or spin-triplet SC disallow Cooper pairs between $k_z$ and $-k_z$ states \cite{Mazin1997,Sigrist2005}. We can also eliminate AFM SF as a candidate mechanism in \KtMA\ via the lack of a resonant-spin excitation in the SC state despite the observation of a SC gap \cite{Zhang2019}. Turning to the upgapped Q1D FSs in \KtCA , this implies that SC must exist on the $\gamma$ sheet. Given the strong FM SF present in \KtCA\ which are known to gap below \Tc\ this points to an interesting scenario \cite{Zhi2015}. While AFM SF cannot pair $k_z$ and $-k_z$ states, FM SF do allow pairing of such states for a gap sign change as occurs in $p_z$-wave orbital symmetry  \cite{Mazin1997}. Furthermore, for FM SF the pairing potential is enhanced for low scattering vectors as found here on the $\gamma$ surface which encompasses the zone center, consistent with the lower $v_F$ found on this sheet \cite{Mazin1997}.  More generally, that both AFM and FM SF exist in \KtCA\ but only the latter responds to SC is highly suggestive of TSC. Thus, our results are consistent with a $p_z$-wave TSC state in \KtCA\ and encourage further work searching for MZMs, potentially pointing to a system which advantageously exhibits a $p_z$-wave state that avoids the singlet-triplet mixing, a highly Q1D crystal habit which may help with device design as well as in isolating such states, and predictions of other intrinsic topological band features \cite{Wu2015,Liu2020,Zhang2019}. 

%Such a scenario which tunes between \textit{e-p} SC in \KtMA\ and SF driven SC in \KtCA\ was suggested recently and is quite interesting showing multiple SC ground states in this family of compounds \cite{Lei2021}. 

%with TSC requiring a sign change of the gap between $k_z$ and $-k_z$ which would counterindicate the attractive SF needed for TSC and singlet SC seeing the converse argument

%Our results offer three conclusions that strengthen the argument for spin-triplet superconductivity in \KtCA : (1) the absence of a spin-gap in the superconducting state of \KtCA\ indicates that the AFM fluctuations are not participating in superconductivity (2) a confirmation of the proposed scenario where \KtMA\ is a \textit{e-p} SC while \KtCA\ is an unconventional SC (3) elimination of AFM fluctuations, the $\gamma$ sheet, and \textit{e-p} coupling from superconductivity in \KtCA\ leaving FM fluctuations as one of the last available candidate pairing interactions. As additional support for this last point, the FM fluctuations are known to be suppressed below \Tc\ and the Q1D FSs are expected to have leading TSC instabilities both of which point to TSC. 

In summary, we show that both \KtCA\ and \KtMA\ exhibit antiferromagnetic spin fluctuations which are consistent with an incipient $\textbf{k}=(0,0,\frac{1}{2})$ ordering vector. Comparing spectra collected above and below their respective \Tc s, we find that while \KtMA\ exhibits a gap, \KtCA\ exhibits no such gap. Furthermore, despite seeing a gap in \KtMA , we observe no evidence of a spin-resonance - the hallmark of spin-driven unconventional superconductivity. Using first principles calculations, we show that these two materials are susceptible to nesting across their Q1D Fermi surfaces consistent with a $\textbf{k}=(0,0,\frac{1}{2})$.  As we observe no gap in the spin-fluctuations of \KtCA , we infer that these Fermi surfaces are not gapped by the superconducting state and that the remaining $\gamma$ sheet, which should favor spin-triplet pairing, must host superconductivity. Furthermore, we rule out the antiferromagnetic and \textit{e-p} coupling superconducting mechanisms in \KtCA , indicating that ferromagnetic fluctuation driven spin-triplet superconductivity is the likely mechanism. As \KtCA\ is a Q1D material, its hosting spin-triplet superconductivity should have exciting implications for topological physics invoking aspects of Kitaev's toy model for Majorana zero-modes.   

\textit{Note:} While this manuscript was in preparation another paper was published (ref.~\onlinecite{Yang2021}) which came to similar conclusions via nuclear magnetic resonance measurements performed on a single crystal sample of \KtCA . These results and ours are quite complementary with the Knight shift providing a more direct measurement of the superconducting state and our measurements and analysis approaching the question of superconductivity in the $A_2TM_3As_3$ family more comprehensively. %Importantly, our work addresses the AFM fluctuations previously observed in \KtCA\ as well as the elucidating the Fermiology behind the observed behaviors. 
%Also gapless spin-fluctuations and luttinger-liquid. 

\begin{acknowledgments}

The authors thank Cristian Batista for helpful conversations pertaining to the significance of the gap in \KtMA . The part of the research that was conducted at ORNL’s High Flux Isotope Reactor was sponsored by the Scientific User Facilities Division, Office of Basic Energy Sciences, US Department of Energy. The research is partly supported by the U.S. Department of Energy (DOE), Office of Science, Basic Energy Sciences (BES), Materials Science and Engineering Division. Work at the University of Missouri is supported by the U.S. DOE, BES, Award No. DE-SC0019114. The part of this work performed at the University of Texas at Dallas is supported by
U.S. Air Force Office of Scientific Research (FA9550-19-1-0037) and National Science Foundation (DMR 1921581). The contribution performed at Air Force Research laboratory was supported by the United States Air Force Office of Scientific Research (AFOSR) LRIR 18RQCOR100 as well as AOARD-MOST Grant Number F4GGA21207H002.

Notice of Copyright This manuscript has been authored by UT-Battelle, LLC under Contract No. DE-AC05-00OR22725 with the U.S. Department of Energy. The United States Government retains and the publisher, by accepting the article for publication, acknowledges that the United States Government retains a non-exclusive, paid-up, irrevocable, world-wide license to publish or reproduce the published form of this manuscript, or allow others to do so, for United States Government purposes. The Department of Energy will provide public access to these results of federally sponsored research in accordance with the DOE Public Access Plan (http://energy.gov/downloads/doe-public-access-plan).

\end{acknowledgments}

% Create the reference section using BibTeX:
%\bibliography{Ref}

%apsrev4-2.bst 2019-01-14 (MD) hand-edited version of apsrev4-1.bst
%Control: key (0)
%Control: author (8) initials jnrlst
%Control: editor formatted (1) identically to author
%Control: production of article title (0) allowed
%Control: page (0) single
%Control: year (1) truncated
%Control: production of eprint (0) enabled
%

\end{document}